\documentclass[preprint,authoryear,12pt]{elsarticle}

\usepackage{amssymb}
\usepackage{amsthm}
\usepackage{array, color}

\newtheorem{theorem}{Theorem}[section]
\newtheorem{lemma}[theorem]{Lemma}
\theoremstyle{remark}
\newtheorem{remark}[theorem]{Remark}
\journal{Computational Statistical and Data Analysis}

\begin{document}

\begin{frontmatter}

%% Title, authors and addresses

%% use the tnoteref command within \title for footnotes;
%% use the tnotetext command for the associated footnote;
%% use the fnref command within \author or \address for footnotes;
%% use the fntext command for the associated footnote;
%% use the corref command within \author for corresponding author footnotes;
%% use the cortext command for the associated footnote;
%% use the ead command for the email address,
%% and the form \ead[url] for the home page:
%%
%% \title{Title\tnoteref{label1}}
%% \tnotetext[label1]{}
%% \author{Name\corref\fnref{label2}}
%% \ead{email address}
%% \ead[url]{home page}
%% \fntext[label2]{}
%% \cortext[cor1]{}
%% \address{Address\fnref{label3}}
%% \fntext[label3]{}

\title{Evaluating the diagnostic powers of variables and their linear combinations  when the gold standard is continuous}

%% use optional labels to link authors explicitly to addresses:
%% \author[label1,label2]{<author name>}
%% \address[label1]{<address>}
%% \address[label2]{<address>}

\author[a,b]{Zhanfeng Wang}
\author[b]{Yuan-chin Ivan Chang \fnref{yc}}
%\ead{ycchang@stat.sinica.edu.tw}
%% \ead[url]{home page}
\fntext[yc]{Corresponding author: ycchang@stat.sinica.edu.tw}

\address[a]{Department of Statistics and Finance, University of Science and Technology of China, Hefei, 230026, China}
\address[b]{Academia Sinica, Taipei, Taiwan, 11529}

\begin{abstract}
The receiver operating characteristic (ROC) curve is a very useful
tool for analyzing the diagnostic/classification power of
instruments/classification schemes as long as a binary-scale gold
standard is available. When the gold standard is continuous and there
is no confirmative threshold, ROC curve becomes less useful.
Hence, there are several extensions proposed for evaluating the diagnostic potential
of variables of interest. However, due to the computational difficulties of these nonparametric based extensions, they are not easy to be used for finding the optimal combination of variables to improve the individual diagnostic power. Therefore, we propose a new measure, which extends the AUC index for identifying variables
with good potential to be used in a diagnostic scheme.  In addition, we propose a threshold gradient descent based algorithm for finding the best linear combination of variables that maximizes this new measure, which is applicable even when the number of variables is huge. The estimate of the proposed index and its asymptotic property are studied. The performance of the proposed method is illustrated using both synthesized and real data sets.

\end{abstract}

\begin{keyword}
%% keywords here, in the form: keyword \sep keyword
ROC curve\sep Area under curve\sep Gold standard \sep Classification
%% MSC codes here, in the form: \MSC code \sep code
%% or \MSC[2008] code \sep code (2000 is the default)

\end{keyword}

\end{frontmatter}

% \linenumbers

%% main text
\section{Introduction}
\label{s:intro}

The ROC curve, founded on a binary gold standard, is one of the most
important tools to measure the diagnostic power of a variable or
classifier, and its importance has been intensively studied by many
authors, which can easily be found in the literature and textbooks
such as \cite{Pepe2003} and \cite{krzanowski-hand2009}.
Moreover, when the number of variables is huge, many algorithms have been
proposed for finding the best combination of variables to increase the individual classification accuracy
(\cite{Su1993}, \cite{Pepe2003}, \cite{Ma2005}, and \cite{Wang-etal07}).
However, in many classification or diagnostic problems, the professed binary gold standard is essentially derived from a continuous-valued variable.
%for example, \cite{pfeiffer05} pointed out that the gold standard may not even exist in many diseases where the work by \cite{choi-et.al-06} was used as a representative of such a situation.
%
If there is no such confirmative threshold for the continuous gold
standard, then the evaluation of variables/classifiers according to
the ROC curve based analysis may vary as the choices of thresholds
change and therefore becomes less informative.
For example, glycosylated hemoglobin is usually used as a primary
diabetic control index, and is originally measured as a
continuous-valued variable. Health institutes, such as the World
Health Organization and National Institutes of Health (NIH), suggest
a cutting point for it  based on current findings for diabetic
diagnosis and control.  Once its cutting point is fixed, then the
association between the variables of interests, such as new drugs,
and this binary-scale standard can be evaluated using some ROC curve
related analysis methods. However, as advances are made in science
and medicine about this disease, this criterion will be re-evaluated
and revised as necessary. Then, the performance evaluation of
variables/classifiers may vary as the binary-recoding scheme is
changed.
%This kind of performances measures, depending on the choice of cutting point of the underlying continuous gold standard, is of course an unwanted character.
It is clear that an unwarranted performance measure may result in
misleading conclusions and may require re-evaluation of all the
available diagnostic methods again every time a new standard is
proposed. Hence, a measure that directly connects to the continuous
gold standard is always preferred, which motivates our study of a
new measure when the gold standard is continuous. Our goal in this
paper  is to find a robust measure, which is not affected by the
choice of cutting point of a gold standard or how the binary outcome
is derived from a continuous gold standard.

%However, this remarkable tool highly relies on the availability of a binary gold
%standard. If the so-called binary gold standard is not available or
%just a surrogate of a continuous-valued variable, the ROC curve will
%suffer from the stability of the underlying binary coding scheme
%that recodes the continuous gold standard to a binary label.

Although there are a lot of reports about the ROC curve, there is
still a lack of study when the gold standard is not binary
\citep{krzanowski-hand2009}.
%Some analysis of ROC curve with continuous gold standard and its related measures have been reported in the literature. For example,
In \cite{henkelman90}, they proposed a maximum likelihood method
under ordinal scale gold standard. Recently, \cite{Zhou-etal05},
\cite{choi-et.al-06}, and \cite{wang-turnbull-et-al07} considered
the ROC curve estimation problems based on some nonparametric and
Bayesian approaches, when there is no gold standard. In addition,
some ROC-type analysis without a binary gold standard has been
considered in \cite{Obuchowski2005} and \cite{Obuchowski2006}, where
a nonparametric method is used to construct a new measure, and many
other applications with continuous gold standard are discussed.
However, these approaches, due to computational issue, are not easy to apply to
the case that the optimal combination of variables is of interest; especially when the
number of variables is large as in modern biological/genetic related studies (\cite{waikar09}).

In this paper, an extension of the AUC-type measure is proposed, which is independent of
the choice of threshold of the continuous gold standard, and algorithms for finding the best linear
combination of variables that maximizes the proposed measure are studied. Under
the joint multivariate normality assumption, the algorithm for the
linear combination can be founded using the LARS method. When this
joint normality assumption is violated, we propose a threshold
gradient descent based method (TGDM) to find the optimal linear
combination. Thus, our algorithms also inherit the nice properties
of LARS and TGDM when dealing with the high dimensional and variable
selection problems. Numerical studies are conducted to evaluate the
performances of the proposed methods with different ranges of
cutting points using both synthesized and real data sets.
The estimate of this novel measure and its asymptotic properties are also presented.

In the next section, we first present a novel measure for evaluating
the diagnostic potential of individual variables and then an
estimate of this measure.  The algorithms for finding the best
linear combination are discussed in Section 3. Numerical results
based on the synthesized data and some real examples follow. A
summary and conclusions are given in Section 4. The technical
details are presented in Appendix.

\section{An AUC-type Measure with a Continuous Gold Standard}

%In real applications, there are usually more than one variables of interest are measured. Therefore, besides, the measure the association between individual variables and the gold standard, we also provide an algorithm to find the best linear combination of variables, which maximizes the value of the proposed measure.

%We know that gold standard in some diagnostic test is not binary and is a continuous output value, then general ROC curve for two-group classification problem is not applied to this situation. In this paper, we propose a ROC-type index to measure the performance of marker to diagnose the studied disease and identify the discriminant markers. Furthermore, we also obtain consistent properties of the ROC measure index. Generally, when many markers simultaneously are measured for studied subject, the combination of markers performs better to associate with gold standard than each single marker. So an optimal linear coefficient of markers associated with gold standard is presented in this paper under the situation of markers and gold standard following a joint multivariate normal distribution. Moreover, when joint distribution of markers and gold standard is unknown, a threshold gradient descend method (TGDM) is proposed to numerically solve the ``optimal'' linear combination coefficient.

%\subsection{An AUC-type Measure Based on A Continuous Gold Standard}
%\setcounter{equation}{00}
Before introducing a novel AUC-type measure based on a continuous
gold standard, we first fix the notation and briefly review the
definition of the ROC curve and its related measures. Let $Z$ and
$Y$ be two continuous real-valued random variables, where $Z$
denotes the gold standard and $Y$ is a variable of interest with
diagnostic potential to be measured. Then, for example, $Z$ is a
primary index for measuring a disease and $Y$ is some other measure
of subjects that is related to the disease of interest. In some
medical diagnostics, the primary index is difficult to measure, and
we are usually looking for variables that are strongly associated
with $Z$ and easy to measure, to be used as surrogates.  That is why
we need to evaluate the ``level of association'' of $Y$ to $Z$.
Likewise, in some bioinformatical studies, in order to develop new
treatments, we would like to identify any strong associations
between some genomic related factors $Y$ to the continuous gold
standard $Z$.
Suppose that there is an unambiguous threshold $c$ of $Z$ that can
be used to classify subjects into two subgroups, and assume further
that subjects with $Z > c$ are classified as diseased, and otherwise
as members of the control group. Then the ROC curve, for such a
given $c$, is defined as $ROC(t) \equiv S_D(S^{-1}_C(t))$, where
$S_D(t)=P(Y >t|Z >c)$ and $S_C(t) = P(Y > t | Z \leq c)$, and the
AUC of variable $Y$ is defined as
\begin{eqnarray}\label{auc_c}
        AUC(c)=P(Y^{+}_c>Y^{-}_c)%={\rm{pr}}(y_1>y_2|z_1>c,z_2<c),
\end{eqnarray}
where random variables $Y^{+}_c$ and $Y^{-}_c$ respectively denote
the $Y$-value of subjects of the disease and non-disease groups with
density functions $f(y|Z>c)$ and $f(y|Z<c)$. That is, $Y^{+}_c$ and
$Y^{-}_c$ are random variables for the sub-populations defined by
$\{Z >c\}$ and $\{Z \leq c\}$, respectively.
It is clear that the $AUC(c)$ defined in (\ref{auc_c}) is a function
of $c$, which will change as the threshold $c$ of $Z$ varies. Hence,
when the threshold is dubious, using $AUC(c)$ as a measure may
misjudge the diagnostic power of $Y$ or the level of association
between $Y$ and $Z$.
%Therefore, how to define a robust measure of diagnostic power of $Y$ when $Z$ is continuous and no confirmative threshold available is an important issue.

Let $f_c(t)$ be a probability density function defined on the range
of possible values of $c$, then $AUC_I$ is defined as
\begin{eqnarray}\label{auc_I}
AUC_I &\equiv \int AUC(t) f_c(t) dt.
\end{eqnarray}
Hence, by its definition, the proposed $AUC_I$ is independent of the
choice of cutting point for the continuous gold standard,
{and any monotonic transformation of $Y$ as well.}
This kind of threshold independent property is also one of the important
properties of the ROC curve and AUC when used as measures of
diagnostic performance.
Since $AUC_I$ is defined as an integration of $AUC(c)$ over the
range of possible cutting points with respect to a weight function
$f_c(t)$, the support of $f_c(t)$ should be chosen as a subset of
the support of the density of $Z$. Moreover, we can use $f_c(t)$ to
put different weights on all possible cutting points of $Z$ if there
is some information about the possible cutting point. If $Z$ is an
ordinal discrete variable, then there are only countable cutting
points, and $f_c(t)$ can be chosen as a probability mass function of
all possible cutting points, and the integration of (\ref{auc_I})
becomes
\begin{eqnarray}\label{auc_s}
    AUC_I&=\sum_{t_i \in C} AUC(t_i) f_c(t_i),
\end{eqnarray}
where $C$ is a set of all possible cutting points. In particular,
when $Z$ is binary, we can let $f_c(t)$ be a degenerated probability
density, then $AUC_I$ is the same as the original AUC.

%Note that using density function $f_c(\cdot)$ is to guarantee that the integration $AUC_I$ has finite value. Two available choices for $f_c$ are $f_c(t)=f_z(t)$ and $f_c(t)=1/a$ in a bound interval with length $a$. We can use index $AUC_I$ to measure power of marker and to identify the discriminant markers.

\subsection{Estimate of ${AUC}_I$}
Let random variables $(Y_i, Z_i)$ denote a pair of measures from
subject $i$, for $i \geq 1$. Suppose that
$\{(y_i,z_i),~i=1,\ldots,n\}$ are $n$ independent observed values of
random variables $(Y_i, Z_i)$, $i=1, \cdots, n$. For a given cutting
point $c$, a subject $i$, $i=1, \ldots, n$, is assigned as a
``case'' if $z_i>c$ and otherwise labeled as a ``control''. That is,
for a given $c$, we divide the observed subjects into two groups;
let $S_1(c)$ and $S_0(c)$ be the case and control groups with sample
sizes $n_1$ and $n_0$, respectively.  It is obvious that these
assignments depend on the choice of $c$. Then for a fixed $c$, the
empirical estimate of $AUC(c)$ is defined as
%For gold standard $Z$ and marker $Y$, we have observed data set $\{(y_i,z_i):~i=1,\cdots,n\}$, and Then the $AUC(c)$ can be estimated with
\begin{eqnarray}\label{AUChat}
    \hat A(c)=\frac{1}{n_0 \, n_1}\sum_{i\in S_1(c); \, j\in S_0(c)} \psi(y_i - y_j),
\end{eqnarray}
where $\psi(u)=1$, if $u>0$; $=0.5$, if $u=0$ and $=0$ if $u<0$. (It
is easy to see that  $\hat A(c)$ does not exist, either
$c>\max\{z_i,i=1,\cdots,n\}$ or $c<\min\{z_i,i=1,\cdots,n\}$, since
for these two cases, we have either $n_1=0$ or $n_0=0$. Therefore,
in this paper, we assume $\hat A(c)=0.5$ when either one of the
cases occurs.)

If the whole support of $Z$ is considered as a possible range of
cutting points, then a natural estimate of $AUC_I$ can be defined as
\begin{eqnarray}\label{gauc_est}
    \hat A_{I}=\int \hat A(t) d \hat F_c(t),
\end{eqnarray}
where $\hat F_c(t)$ is the empirical estimate of the cumulative
distribution function of $Z$ based on $\{z_1, \ldots, z_n\}$.
However, in practice, it is rare to choose cutting points at ranges
near the two ends of the distribution of $Z$.  Thus, instead of the
whole range of $Z$, we might explicitly define a weight function
$f_c(t)$ on a particular critical range. Below, we demonstrate three
possible choices: (1) a uniform distribution over the range of
$(-\hat\sigma, +\hat\sigma)$, where $\hat\sigma$ is an empirical
standard deviation of $Z$, say $f_1(t)$; (2) a normal density with
sample mean $\hat \mu$ and standard deviation $\hat\sigma$ based on
the observed values of $Z$, say $f_2(t)$; or (3) using a kernel
density estimate, say $f_3(t)$, to approximate the marginal density
of $Z$.
For different weight functions $f_j(t)$, $j=1, 2, 3$, the estimate
of $AUC_I$ is denoted as
\begin{eqnarray}\label{gauc_e}
    \hat A_{Ij}=\int \hat A(t) f_j(t) dt.
\end{eqnarray}
It is clear that our method can be extended to other reasonable
choices of weight functions. The theorem below states the strongly
consistent property of $\hat A_{Ij}$ for all $j$.
%Taking a closer look to the expression (\ref{gauc_e}), we see that the $\hat A(c)$ does not exist by using the definition of expression (\ref{AUChat}) when $c>\max\{z_i,i=1,\cdots,n\}$ or $c<\min\{z_i,i=1,\cdots,n\}$, since $n_1=0$ or $n_0=0$. Fortunately, we know minimum value of AUC is 0.5 which indicates that subjects are randomly classified into diseased or non-disease
%groups by using marker $Y$. Therefore, in this paper we define $\hat A(c)=0.5$ when $c>\max\{z_i,i=1,\cdots, \}$ or $c<\min\{z_i,i=1,\cdots,n\}$.
\begin{theorem}\label{thm1-gauc}
    Let $(Y\in R^1, Z\in R^1)$ be a pair of random variables with uniformly continuous marginal densities.  Assume that
    $\{(y_1,~z_1),\ldots,(y_n,~z_n)\}$ are $n$ observations of the independent and identically distributed random sample
    $(Y_i, Z_i)$, $i=1, \ldots, n$. Assume further that $Z$ is the continuous gold standard.
    Then for a given $f_c(t)=f_j(t)$, $j=1,~2,~3$, with probability one, $\hat A_{Ij}-AUC_{Ij} \rightarrow 0 \hbox{~as~} n \rightarrow \infty$,
    where $\hat A_{Ij}$ and $AUC_{Ij}$  are defined as in (\ref{gauc_e}) and(\ref{auc_I}), respectively, with corresponding $f_c(t)=f_j(t)$.
\end{theorem}
%
%\begin{proof}[ of Theorem \ref{thm1-gauc}]
\noindent{\bf Proof of Theorem \ref{thm1-gauc}} Since bounded function $\hat A(c)$
converges almost surely to ${AUC}(c)$ for all given $c$ and $f_c(t)$ is also
bounded density function, the proof of Theorem \ref{thm1-gauc} follows from
the dominated convergence theorem.
%\end{proof}

It is difficult to have an explicit form for the variance of $\hat
A_{Ij}$ due to its integral form.  Thus, a bootstrap estimate of the
variance of $\hat A_{Ij}$ is used and denoted as $V(\hat A_{Ij})$.
A similar idea is employed in \cite{Obuchowski2006}.
{
\begin{remark}
  Note that the method for calculating (\ref{gauc_e}) may depend on the choice of weight function. If the empirical density of the gold standard is used, then the computation of it is straightforward; if a kernel density of the gold standard is used, then a numerical integration method is required. However, in all cases the computation of it are easy since it is an one-dimensional density.
\end{remark}
}
\section{Linear combination of variables that maximizes $AUC_I$}
%\vskip 10pt
For a classification or diagnostic problem, there are usually many
variables measured from each subject, and it is well known that a
combination of variables can usually improve on the classification
performance of a single variable. This situation motivates us to
study how to find the optimal linear combination of variables that
maximizes the proposed measure $AUC_I$. For classical AUC,
\cite{Su1993} studied the best linear combination under a
multivariate normal distribution assumption. Here we extend their
idea to $AUC_I$. In addition, we also aim to address cases with
huge number of variables, which usually involve some computational issues and will be discussed later in this section.

\subsection{Optimal Linear Combination of Variables Under Joint Normality}
For clarity and convenience, we start with a bivariate normal
distribution case, since the linear combination of variables, for a given vector of coefficients, can be
treated as a single variable.

Let $U=(Y,Z)^{T}$ be a random vector following a bivariate normal
distribution with mean vector $\mu=(\mu_1,\mu_2)^{T}$ and covariance
matrix
\begin{eqnarray}
\Sigma_U=\left(
  \begin{array}{cc}
    \sigma_1^2 & \rho \sigma_1\sigma_2 \\
    \rho \sigma_1\sigma_2 & \sigma_2^2 \\
  \end{array}
\right).\nonumber
\end{eqnarray}
Suppose that $U_i=(Y_i, Z_i)^{T}$, $i=1, 2$, are two independent
random vectors generated from the same distribution of
 $U$. Define
\[Q_i=\exp{\left(-\frac{(U_i-\mu)^{T}\Sigma_U^{-1}(U_i-\mu)}{2}\right)},~ i=1,2.\]
Then for a given $c$,
\begin{eqnarray}\label{eqn1}
{\rm{pr}}(Y_1>Y_2,Z_1>c,Z_2<c)=\int_{-\infty}^{\infty}\int_{-\infty}^{y_1}\int_{c}^{\infty}\int_{-\infty}^{c}
\frac{Q_1 Q_2}{4\pi^2|\Sigma_U|}\, dz_2 dz_1 dy_2 dy_1,
%\exp{\left(-\frac{(U_1-\mu)^{'}\Sigma_1^{-1}(U_1-\mu)}{2}\right)}\nonumber\\
%& \hskip 3cm \exp{\left (-\frac{(U_2-\mu)^{'}\Sigma_1^{-1}(U_2-\mu)}{2} \right)}dz_2 dz_1 dy_2 dy_1.
\end{eqnarray}
where $|\Sigma_U|$ denotes the determinant of matrix $\Sigma_U$.
%
%Under the joint normal assumption of $U$,
The conditional distribution of $Y_j$ given $Z=z_j$ is a normal
distribution with mean
$\tilde{\mu}_j=\mu_1+\sigma_1/\sigma_2\rho(z_j-\mu_2)$ with $j=1,~2$ and variance
$\tilde{\sigma}_1^2=(1-\rho^2)\sigma_1^2$.
%and $Y_2$ conditional on $Z_2=z_2$ has a normal distribution with mean $\tilde{\mu}_2=\mu_1+\sigma_1/\sigma_2\rho(z_2-\mu_2)$ and variance $\tilde{\sigma}_1^2$.
Let $\eta(z_1,z_2)={1}/{(2\pi
\sigma_2^2)}\exp(-((z_1-\mu_2)^2+(z_2-\mu_2)^2)/{(2\sigma_2^2)})$.
Then,  (\ref{eqn1}) can be rewritten as
\begin{eqnarray}\label{eqn2}
&& {\rm{pr}}(Y_1>Y_2,Z_1>c,Z_2<c)\nonumber\\
&&=\int_{c}^{\infty}\int_{-\infty}^{c}\eta(z_1,z_2)\int_{-\infty}^{\infty}\int_{-\infty}^{y_1}
\frac{1}{2\pi \tilde{\sigma}_1^2} \exp{\left(-\frac{(y_1-\tilde{\mu}_1)^2
+(y_2-\tilde{\mu}_2)^2}{2\tilde{\sigma}_1^2}\right)}dy_2 \nonumber\\
&&\hskip 11cm dy_1 dz_2 dz_1 \nonumber\\
%&=\int_{c}^{\infty}\int_{-\infty}^{c}
%\eta(z_1,z_2) E_{\tilde{Y}_1}\Phi(\frac{\tilde{Y}_1-\tilde{\mu}_2}{\tilde{\sigma}_1})dz_2 dz_1, \nonumber \\
&&=\int_{c}^{\infty}\int_{-\infty}^{c}
\eta(z_1,z_2) {\rm{E}}(\Phi(\frac{\tilde{\sigma}_1 V+\tilde{\mu}_1-\tilde{\mu}_2}{\tilde{\sigma}_1})) dz_2 dz_1\nonumber\\
&&=\int_{c}^{\infty}\int_{-\infty}^{c}\eta(z_1,z_2)
{\rm{E}}(\Phi(V+\frac{\rho(z_1-z_2)} {{\sigma}_2(1-\rho^2)^{1/2}}))
dz_2 dz_1,
\end{eqnarray}
where %$\tilde{Y}_1$ follows normal distribution with mean $\tilde{\mu}_1$ and variance $\tilde{\sigma}_1^2$
$V$ is a standard normal random variable and $\Phi$ is the standard
normal cumulated distribution function. {Note that under normality assumption,
$\rho=0$ implies that $Y$ and $Z$ are independent, and it follows from (\ref{eqn2}) $AUC_I=0.5$ in this case.}
%We know that in the classic ROC curve analysis, the $AUC=0.5$ does not imply the ROC curve is lying in the diagonal of the unit square.

Now, suppose that $\tilde{X}=(X_1,\ldots,X_p)^{T}$ is a
$p$-dimensional random vector of measures of a subject, and $Z$ is
the continuous gold standard as before. Suppose $l\in R^p$ and let
$Y=l^{T}\tilde{X}$ be a linear combination of $\tilde{X}$. Assume
further that $\tilde{X}$ follows a multivariate normal
distribution with mean vector $\mu^*$ and covariance matrix
$\Sigma$. Then $Y$ follows a normal distribution with mean
$\mu_1=l^{T}\mu^*$ and variance $\sigma_1^2=l^{T}\Sigma l$. The
correlation coefficient between $Y$ and $Z$ is $\rho=l^{T}{\rm
cov}(\tilde{X},Z)/((l^{T}\Sigma l)^{1/2}\sigma_2)$, where ${\rm
cov}(\tilde{X},Z) = ({\rm cov}(X_1,Z),\ldots,{\rm cov}(X_p,Z))^{T}$.
Then, $AUC_I$ for such a linear combination of $X_i$'s,
$Y=l^{T}\tilde{X}$, is a function of $l$:
\begin{eqnarray}\label{linauc}
AUC_{I}(l)&=&\int {\rm{pr}}(l^{T}\tilde{X}_1>l^{T}\tilde{X}_2|Z_1>t,Z_2<t) f_c(t) dt%\nonumber\\
%&=&\int \frac{{\rm{pr}}(l^{T}\tilde{X}_1>l^{T}\tilde{X}_2,Z_1>t,Z_2<t)}{{\rm{pr}}(Z_1>t,Z_2<t)}f_c(t) dt,
\end{eqnarray}
where $(\tilde{X}_i^{T},Z_i)^{T}$, $i=1,2$, are independent
identically distributed samples of $(\tilde{X}^{T},Z)^{T}$.
Our goal is to find the optimal linear combination of $X_1,\ldots,
X_p$ such that $AUC_I$ is maximized and it is known that AUC is
scale invariant. In order to make the solution identifiable, we
search for an $l_{opt}$ such that $AUC_I(l_{opt}) \geq AUC_I(l)$ for
all possible $l \in R^p$ with $\| l \|=1$.

From (\ref{eqn2}),
\begin{eqnarray}\label{eqn3}
    \frac{\partial }{\partial l}{\rm{E}}\left(\Phi\left(V+\frac{\rho(z_1-z_2)}
{{\sigma}_2(1-\rho^2)^{1/2}}\right)\right)=
{\frac{1}{\sqrt{2}}}\exp\left(-\frac{\rho^2(z_1-z_2)^2}{4\sigma_2^2(1-\rho^2)} \right)
\frac{z_1-z_2}{\sigma_2(1-\rho^2)^{3/2}}\frac{\partial \rho}{\partial l}.
\end{eqnarray}
Therefore,
\begin{eqnarray}\label{eqn4}
    \frac{\partial AUC_{I}(l)}{\partial l}&=&\frac{\partial \rho}{\partial l} \int f_c(t) \int_{t}^{\infty}\int_{-\infty}^{t} \frac{1}{{2}^{3/2}\pi \sigma_2^2} \exp{\left (-\frac{(z_1-\mu_2)^2+(z_2-\mu_2)^2}{2\sigma_2^2}\right)} \nonumber\\
    & &\exp\left(-\frac{\rho^2(z_1-z_2)^2}{4\sigma_2^2(1-\rho^2)} \right) \frac{z_1-z_2} {\sigma_2(1-\rho^2)^{3/2}}\frac{1}{{\rm{pr}}(Z_1>t,Z_2<t)}dz_2 dz_1 dt\nonumber\\
    &=&\frac{\partial \rho}{\partial l}\Delta,
\end{eqnarray}
where $\Delta$ dentes the integration part of the left hand side of
(\ref{eqn4}). Since $\Delta>0$, the equation ${\partial
AUC_I(l)}/{\partial l}=0$ if and only if ${\partial \rho}/{\partial
l}=0$; that is,
\[
    \frac{\partial}{\partial l}\frac{l^{T}{\rm cov}(\tilde{X},Z)}{((l^{T}\Sigma l)^{1/2}\sigma_2)}=0.
\]
It implies that the optimal linear combination coefficient
\begin{eqnarray}\label{eqn5}
    l_{opt}=\Sigma^{-1}{\rm cov}(\tilde{X},Z).
\end{eqnarray}
%Thus, the $AUC_{I}(l_{opt})$ reaches its maximum value of $AUC_{I}(l)$.
%\textcolor{red}{(Why does it imply the solution will achieve the maximum?)}
{Note that, as in \cite{Su1993}, this optimal linear combination coefficient $l_{opt}$ is independent
of $c$, and depends only on the covariance matrix of variables and the covariance between of
variables and the gold standard.}

\subsection{Estimation of the Optimal Linear Combination}
Assume that $\{(\tilde{x}_i,z_i), i=1,\cdots,n\}$ is a set of $n$
independent and identically distributed random samples, where $z_i$
denotes the observed gold standard measures as before, and
$\tilde{x}_i$ is its corresponding $p$-dimensional vector of
observed variable values of subject $i$. Without loss of generality,
we assume that all the components of $\tilde{x}$ and $z$ are
centralized, since we can always centralize the data by subtracting
their sample means, and define
$H=(\tilde{x}_1-\bar{x},\cdots,\tilde{x}_n-\bar{x})^{T}$ as an $n
\times p$ matrix, and
$\tilde{z}=(z_1-\bar{z},\cdots,z_n-\bar{z})^{T}$ as a vector of
length $p$, where $\bar{x}=\sum_{i=1}^n \tilde{x}_i/n$ and
$\bar{z}=\sum_{i=1}^n z_i/n$.
Hence, the estimate of $l_{opt}$ based on a sample of size $n$
following from (\ref{eqn5}) is defined as
\begin{eqnarray}\label{eqn6}
    \hat l= (H^{T}H)^{-1}H^{T}\tilde{z}.
\end{eqnarray}
Similarly to the linear regression problem, it is clear that $\hat
l$ is a strongly consistent estimate of $l_{opt}$ under some
regularity conditions on $\tilde X$ and $Z$.
%when the joint distribution of $\tilde{X}$ and $Z$ satisfy a multivariate normality assumption.
%
Define
\begin{eqnarray}\label{AUC_e}
    \hat A(c, l)=\frac{1}{n_1 n_0}\sum_{i\in S_1(c);j\in S_0(c)}\psi(l^{T}\tilde{x}_i-l^{T}\tilde{x}_j).
\end{eqnarray}
Then
\begin{eqnarray}\label{elinauc}
        \hat A_{I}(l)=\int \hat A(t, l) f_c(t) dt
\end{eqnarray}
is an estimate of $AUC_{I}(l)$. It is easy to see that for given $t$,
$\hat A(t, l)$ convergenes to   $\hat A(t, l)$ uniformly with respect to $l$.
Hence, using the dominated convergence theorem, it is shown that $\hat A_{I}(\hat l)$ is a strongly
consistent estimate of $AUC_{I}(l_{opt})$ and the details are
omitted here. This result is stated as a theorem below:
\begin{theorem}\label{thm2-gauc}
    Suppose that the joint distribution of $\tilde{X} \in R^p, Z \in R^1$ follows a multivariate normal distribution, where $Z$ is the continuous gold standard, and $\tilde{X}$ denotes the $p$-dimensional vector of variables. Let $\{(\tilde{X}_1,~Z_1),\cdots,(\tilde{X}_n,~Z_n)\}$ be independent and identically distributed samples of size $n$. Then for a given density $f_c(t)$, with probability one,
\begin{eqnarray}
    \hat A_{I}(\hat l)-AUC_{I}(l_{opt})\longrightarrow 0, \hbox{~as~} n \rightarrow \infty,\nonumber
\end{eqnarray}
where $AUC_{I}(l_{opt})$ and $\hat A_{I}(\hat l)$ are defined as in
(\ref{linauc}) and  (\ref{elinauc}) with $l=l_{opt}$ and $\hat l$,
respectively.
\end{theorem}

Equation (\ref{eqn6})  provides a neat solution for the best linear
combination of variables under a joint multivariate normality
assumption.  However, it can be seen from (\ref{eqn6}) that the
calculation of $\hat l$ relies on the computation of an inverse
matrix. Thus, when the number of variables is large, the direct
calculation of $\hat l$ using (\ref{eqn6}) becomes numerically
unstable. The situation is worse, when the sample size is relatively
small compared to the number of variables. So, we need  an
alternative numerical approach that can handle  problems with large
$p$ to overcome this obstacle.

Again, from (\ref{eqn6}), we find  that the estimate $\hat l$ can be
viewed as a least square estimate of $l$ in the linear regression
model below:
\begin{eqnarray}\label{eqn7}
     \tilde{z}=H l+e,
\end{eqnarray}
where  $e$ is an $n$-dimensional vector of random error. When $p$ is
small, then the solution can be obtained easily as in regression
problems. When $p$ is large, then we can apply the least angle
regression shrinkage (LARS) method \citep{Efron2004} to (\ref{eqn7})
to obtain an estimate of $l$.  Since this is the same as applying
LARS in a regression setup, the properties of LARS are therefore
inherited.  With the assistance of LARS, the proposed measure can be
applied to evaluate linear combinations of lengthy variables.  The
variable selection scheme will follow from LARS as it is used in
regression models. However, when the normality assumption is
violated or the normal approximation to the joint distribution is
not adequate, the empirical results show that the $l_{opt}$ defined
in (\ref{eqn5}) is not a good solution. Thus, an alternative
algorithm, which does not rely on the normality assumption, is
required and developed below.
\begin{remark}
Since the properties of applying LARS to find the linear combination
of variables are the same as those in linear regression. We omit the
details of applying LARS under the normality assumption. Instead, we
focus on the case without a normality assumption.
\end{remark}

%\subsection{Computation of $AUC_{Is}$ Without Normality Assumption}
\subsection{When the Joint Distribution is Unknown}
%\vskip 10pt
%When the joint distribution of $\tilde{X}$ and $Z$  is unknown, or only limited information of it is available such that the normal approximation is not appropriate, we propose an algorithm to find the ``optimal'' linear coefficient of variables, when the normality assumption is not fulfilled.
As before, let's start with a one-dimensional case, and the case
with a linear combination of variables will follow easily as an
extension.

%is not continuously differentiable due to the step function $\psi(\cdot)$ in (\ref{AUChat}), and
Similarly to the methods used in \cite{Ma2005}, and
\cite{Wang-etal07}, we first use a sigmoid function
$S(t)=1/(1+\exp(-t))$ to approximate $\psi(\cdot)$ in equation
(\ref{gauc_est}). Thus, a smooth estimate of $AUC_I$ is defined as
\begin{eqnarray}\label{sauc_I}
    \hat A_{Is}=\int \frac{1}{n_1 n_0}\sum_{i\in S_1(t);j\in
    S_0(t)}S\left (\frac{y_i-y_j}{h}\right )f_c(t) dt.
\end{eqnarray}
It follows from the results in density estimation literature that
for a sufficiently small window width $h$, $S((y-x)/h) \approx
\psi((y-x)$, which implies the following asymptotic properties of
$\hat A_{Is}$:
\begin{theorem}\label{thm3-gauc}
    Assume that $\{(y_1,~z_1),\cdots,(y_n,~z_n)\}$ are n independent and identically distributed samples of $(Y \in R^1, Z\in R^1)$, where $Z$ denotes a continuous gold standard. Denote the marginal densities of $Y$ and $Z$ by $f_{Y}$ and $f_{Z}$, respectively. Let $F(z|y)$ be conditional cumulative function of $Z$ given $Y=y$. Suppose that $f_{Y}$ and $f_{Z}$ are larger than 0 and bounded. Assume both $f_{Y}(\cdot)$ and $F(z|\cdot)$ are uniformly continuous. Then for a given probability density $f_c(t)$ with $h=O(n^{-\alpha})$, $1/5<\alpha<1/2$,
    \begin{eqnarray}
        \hat A_{Is}-AUC_{I}\rightarrow 0 \hbox{~~ almost surely} \hbox{~as~} n \rightarrow \infty,\nonumber
    \end{eqnarray}
    where $AUC_{I}$  and $\hat A_{Is}$ are defined in (\ref{auc_I}) and  (\ref{sauc_I}),  respectively.
\end{theorem}
(The proof of Theorem \ref{thm3-gauc} relies on some classical
results of density approximation theory. The details are given in Appendix A.)

As before, we replace $y$ in (\ref{sauc_I}) with $l^T\tilde x$, then
we have the smooth estimate of $AUC_I(l)$ below:
\begin{eqnarray}\label{slinauc}
    \hat A_{Is}(l) =\int \frac{1}{n_1 n_0}\sum_{i\in S_1(t);j\in
    S_0(t)}S\left (\frac{l^{T}\tilde{x}_i-l^{T}\tilde{x}_j}{h}\right )f_c(t) dt.
\end{eqnarray}
The asymptotic property of $\hat A_{Is}(l)$  follows easily from
Theorem \ref{thm3-gauc}, and is summarized as the following theorem
without proof.
\begin{theorem}\label{thm4-gauc}
Suppose that $\{(\tilde{x}_1,~z_1),\cdots,(\tilde{x}_n,~z_n)\}$ are
$n$ independent and identically distributed samples of $(\tilde{X}
\in R^p, Z\in R^1)$, where $Z$ denotes the continuous gold standard,
and $\tilde{X}$ is a vector of corresponding variables. Let $f_c(t)$
be a probability density. Assume that for a given constant vector
$l\in R^p$,  the conditions of Theorem (\ref{thm3-gauc}) holds for
$Y=l^{T}\tilde{X}$ and $Z$. Then for $h=O(n^{-\alpha})$ with
$1/5<\alpha<1/2$,
\begin{eqnarray}
    \hat A_{Is}(l)-AUC_{I}(l)\rightarrow 0 \hbox{ almost surely} \hbox{ as } n \rightarrow \infty,\nonumber
\end{eqnarray}
where $AUC_{I}(l)$ and $\hat A_{Is}(l)$ are defined in
(\ref{linauc}) and (\ref{slinauc}), respectively.
\end{theorem}
\begin{remark}
  We only need to estimate the density function of the linear combination $l^{T}\tilde{X} \in R^1$, hence the choice of $h$ does not depend on the length of total variables $p$. Thus, the density estimation part of the proposed algorithm will not suffer from the curse of dimensionality.
\end{remark}

Following Theorem \ref{thm3-gauc}, we apply the threshold gradient
descent method (TGDM) of \cite{Friedman2004} to find the best linear
combination, $\hat{l}$ which maximizes $\hat A_{Is}(l)$. That is, to
find a solution
\begin{eqnarray}\label{lhat}
    \hat{l}=\mbox{argmax}_{l}\hat A_{Is}(l).
\end{eqnarray}
From equation (\ref{slinauc}), we know that $AUC_{Is}$ is also scale
invariant as is AUC. That is, $\hat A_{Is}(l)$ with window width $h$
will equals to $\hat A_{Is}(kl)$ with $h=kh$ for a positive constant
$k$. Hence, an anchor variable is needed such that the solution of
(\ref{lhat}) is unique.

\vskip 10pt \noindent\textbf{TGDM Based
Algorithm} Let $\{(\tilde{x}_1, z_1),\cdots,(\tilde{x}_n, z_n)\}$ be
a set of random samples of size $n$, which satisfies the assumption
of Theorem \ref{thm4-gauc}.  Define $s=(s_1,\cdots,s_{p})^{T}$ as a
$p$-dimensional vector with $s_i=1$, if the corresponding empirical
$AUC_I$ of the $i$th variable is greater than 0.5; otherwise set
$s_i=-1$. Let $\beta_i$ be a $p$-dimensional vector where only the
$i$th component equals $s_i$ and $0$ otherwise. Define $R_i=\hat
A_{Is}(\beta_i)$, then choose the variable with the maximum $R_i$
value as the anchor variable. In the following algorithm, we assume
that $R_1>R_i$, for $i=2, \ldots, p$ without loss of generality. Let
notation $\hat l_{i}$ denote the $i$th component of $\hat l$, then
$\hat l_{1}$ is the coefficient of the anchor variable. In order to
make the coefficients identifiable, we set $\|\hat{l}_{1}\|=1$.
Following the notations defined above, a TGDM-based algorithm for finding the best linear combination of variables that maximizes $AUC_{Is}$ is stated below:\\
\vskip 5pt \noindent\textbf{Algorithm:}
\begin{itemize}
\vspace{-3pt}
\item[](0) Initial stage:  Let $r=0$ and  choose a threshold parameter $\tau$.
Set $l^{(0)}=(s_1,0,\cdots,0)^{T}$.
\item[](1) Given $l=l^{(r)}$, calculate the derivative of the smoothed estimate $\hat A_{Is}(l)$
with respect to linear coefficient $l$,
$d(l^{(r)})=(d_1(l^{(r)}),\cdots,d_p(l^{(r)}))^{T}=\partial\hat
A_{Is}(l)/\partial l|_{l=l^{(r)}}$.
\item[](2) Use the threshold gradient descent method to calculate $l=l_0^{(r+1)}$;\\ that is,
    $l_0^{(r+1)}=l^{(r)}+\delta~ t(\tau,l^{(r)})~ d(l^{(r)})$ for some $\delta>0$, where
    $t(\tau,l^{(r)})$ is an indicator vector $$I\left(d(l^{(r)}) > \tau~ {\rm {max}} \{d_1(l^{(r)}),\cdots,d_p(l^{(r)})\}\right).$$
\item[](3) Find the optimal $\delta^*\!\!=\!\!{\rm {argmax}}_{\delta>0} \hat A_{Is}(l_0^{(r+1)})$ with $l_0^{(r+1)}\!\!=l^{(r)}+ \delta t(\tau,l^{(r)})~ d(l^{(r)})$, and update $l^{(r+1)}=l^{(r)}+\delta^* t(\tau,l^{(r)})~ d(l^{(r)})$.
\item[](4) Repeat steps (1)-(4) until $\hat A_{Is}(l^{(r+1)})$ converges.
\end{itemize}
\begin{remark}
    The initial value of $l$ is chosen as $(s_1,0,\cdots,0)^{T}$, since the first component of $l$ corresponds to the selected anchor variable.
    In Step (2), we update $l^{(r)}$ along the direction $t(\tau,l^{(r)})~ d(l^{(r)})$, where the number of  nonzero components is decided by the threshold parameter $\tau$, and by the definition of $t(\tau,l^{(r)})$, the locations of nonzero components of $t(\tau,l^{(r)})$ are determined by the elements of gradient $d(l^{(r)})$.
    Step (3) is to find a suitable step size $\delta^*$ along the direction of Step (2), then update the linear coefficients of variables.
    The criterion of convergence of Step (4) has to be predetermined.\\
    (The software used in this paper (GoldAUC) is available at\\ {http://idv.sinica.edu.tw/ycchang/software.html}).
\end{remark}

\section{Numerical studies}
%\setcounter{equation}{00}
%For comparison purpose, we briefly described the approach \cite{Obuchowski2006}.
In numerical studies, we calculate the proposed measures $\hat
A_{Ij}$, $j=1, 2, 3$, corresponding to 3 different $f_c(t)$ as
defined before. Since the correlation coefficient is a basic
statistic to measure the association between two continuous
variables, we therefore include it in our experimental studies. We
also compare the performances of our methods with that of
Obuchowski's (2006) method (page 485, Equation (9)) described below:
%When the gold standard is a continuous variable,  \cite{Obuchowski2006} proposed a ROC-type statistic $\hat\theta$ based on some nonparametric method as follows:
\begin{eqnarray}\label{theta}
    \hat\theta=\frac{1}{n(n-1)}\sum_{i=1}^n\sum_{j=1}^n
    \psi^{'}(y_i,z_i,y_j,z_j),
\end{eqnarray}
where $i\not=j$,
\begin{eqnarray}
  \psi^{'}(y_i,z_i,y_j,z_j) &=& 1  \rm{~~~~ if~~ } y_i>y_j \rm{~~ and~~ } z_i>z_j, \rm{~~ or ~~} y_i<y_j \rm{~~ and ~~}
  z_i<z_j; \nonumber \\
  &=&0.5 \rm{~~ if~~ } y_i=y_j \rm{~~ or~~ } z_i=z_j; \nonumber \\
  &=&0 \rm{~~~~ otherwise.} \nonumber
\end{eqnarray}

% Set up of experimental study
The sample sizes used in our numerical studies are $n=50$ and $100$.
The window width for the kernel estimate in $\hat A_{I3}$ is equal
to $n^{1/5}$. The bootstrap sample size for estimating the variance
of each case is 200, and there are 100 replicates for each
simulation setup.
For the first experimental study, the data are generated from
bivariate normal distributions with means $\mu_1=\mu_2=1.0$,
standard deviations $(\sigma_1, \sigma_2)$ equal to $(1.0,1.0)$,
$(1.0, 2.0)$, $(2.0,1.0)$ and $(2.0,2.0)$, and correlation
coefficients equal to $\rho=0.0$, $0.25$, $0.5$, $0.75$, and $1.0$.
Let $\hat\mu$ and $\hat\sigma^2$ denote the sample mean and variance
of $z$.
{
%\begin{remark}
  As in the classical ROC curve analysis, when a variable with no diagnostic power, then its corresponding
  ROC curve will be the 45 degree diagonal line of the unit square.
If this case holds for all possible cutting points, then it implies that $AUC_I=0.5$.
So, we use 0.5 as the value of the null hypothesis in our numerical study.
%\end{remark}
}
Table \ref{tab1} shows five statistics for different
simulation setups: correlation coefficient of two variables $\hat
\rho$,  $\hat A_{Ij}$, $j=1, 2,3$ with corresponding $f_c(t)$'s, and
$\hat \theta$ from \cite{Obuchowski2006}.
Figure 1 is a plot of statistics $\hat \rho^2/V(\hat \rho)$, $(\hat
A_{Ij}-0.5)^2/V(\hat A_{Ij})$ for all $j$'s, and $(\hat
\theta-0.5)^2/V(\hat \theta)$ versus $\rho$, where $V(\hat \rho)$
and $V(\hat\theta)$ are the bootstrap estimates of variances of
$\hat \rho$ and $\hat\theta$, respectively.

\begin{figure}
\vskip 10pt \centering
\includegraphics[width=5in]{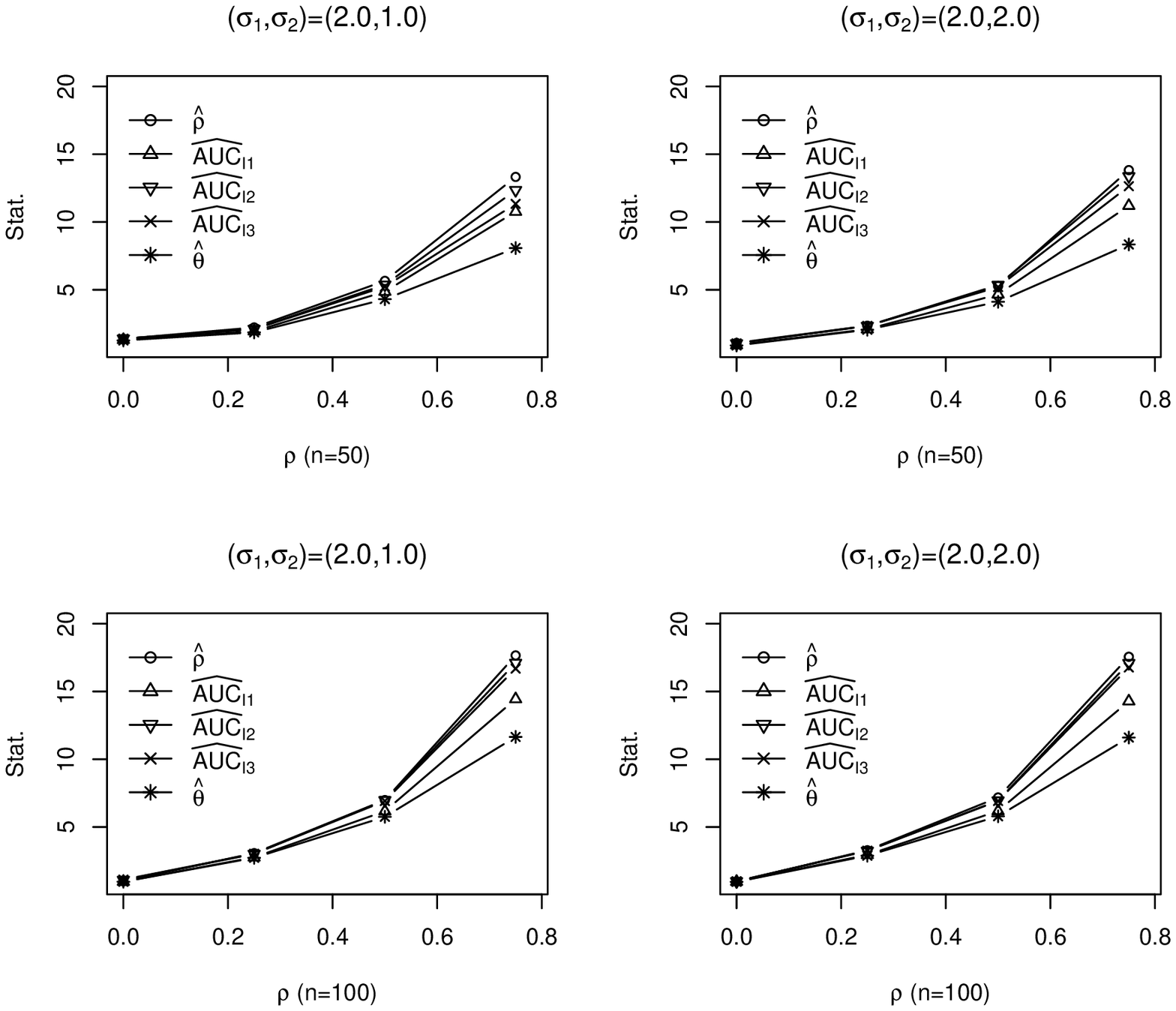}
\caption{Comparison of five measures: $\hat \rho^2/V(\hat \rho)$,
$(\hat A_{Ij}-0.5)^2/V(\hat A_{Ij}),~j=1,~2,~3$, and $(\hat
\theta-0.5)^2/V(\hat \theta)$, where $(Y, Z)$ follow bivariate
normal distributions with means $\mu_1=\mu_2=1.0$, with different
standard deviations $\sigma_1,~\sigma_2$ and correlation
coefficients $\rho$.} \label{fig1}
\end{figure}

When the joint distribution of two variables follows a bivariate
normal distribution, the correlation coefficient is a natural
statistic to describe the association between the two variables. In
our study, all five measures increase as the true correlation
coefficient $\rho$ increases, which suggests that all measures catch
the linear association between variable $Y$ and the gold standard
$Z$ as expected. In fact, $\hat A_{Ij}$ and $\hat\theta$ are very
close to their true values 0.5 and 1.0, when $\rho$ are equal to 0.0
and 1.0, respectively.
%
%It indicates that indexes $\hat A_{Ij}$ and $\hat\theta$ are reasonable for marker having no/perfect association with gold standard.
%
In addition, Figure \ref{fig1} shows that the values of $\hat
\rho^2/V(\hat \rho)$ and $(\hat A_{Ij}-0.5)^2/V(\hat A_{Ij})$, $j=1,
2,3$, are larger than those of $(\hat\theta-0.5)^2/V(\hat \theta)$ under current simulation set up.

\begin{table*}[]
\tabcolsep=1.5pt \fontsize{7}{9}\selectfont \caption{Comparison of
five measure indexes: $\hat \rho$, $\hat A_{Ij},~j=1,~2,~3$, and
$\hat \theta$, where the marker and gold standard, $(y,~z)$, follow
multi-variate normal distribution with means $\mu_1=\mu_2=1.0$,
with different standard deviations $\sigma_1,~\sigma_2$ and distinct correlation coefficients $\rho$.}\label{tab1}{%
 \begin{tabular*}{\textwidth}{@{}c@{\extracolsep{\fill}}c@{\extracolsep{\fill}}c@{\extracolsep{\fill}}c
 @{\extracolsep{\fill}}c@{\extracolsep{\fill}}c@{\extracolsep{\fill}}c@{\extracolsep{\fill}}c}
 %@{\extracolsep{\fill}}c@{\extracolsep{\fill}}c@{}}
%\begin{tabular*}{cccc cccc}
 \hline
 % after \\: \hline or \cline{col1-col2} \cline{col3-col4} ...
$n$&$(\sigma_1,~\sigma_2)$ &Method&0.0&  0.25& 0.5& 0.75& 1.0\\

50&(1.0,~1.0)&$\hat\rho$&0.105(0.076,~0.140)$^{*}$&0.252(0.118,~0.130)&0.511(0.088,~0.103)&0.747(0.067,~0.064)&1.000(0.000,~0.000)\\
&&$\hat A_{I1}$&0.505(0.064,~0.073)&0.621(0.063,~0.069)&0.746(0.053,~0.058)&0.866(0.040,~0.038)&1.000(0.000,~0.000)\\
&&$\hat A_{I2}$&0.501(0.065,~0.067)&0.616(0.062,~0.063)&0.743(0.046,~0.052)&0.856(0.035,~0.033)&0.979(0.010,~0.013)\\
&&$\hat A_{I3}$&0.498(0.065,~0.066)&0.611(0.062,~0.062)&0.737(0.045,~0.051)&0.846(0.037,~0.034)&0.968(0.011,~0.015)\\
&&$\hat \theta$&0.504(0.044,~0.049)&0.583(0.044,~0.048)&0.673(0.038,~0.044)&0.771(0.036,~0.035)&1.000(0.000,~0.004)\\ \\
%\cline{3-8}
&(1.0,~2.0)&$\hat\rho$&0.106(0.073,~0.136)&0.263(0.118,~0.131)&0.477(0.099,~0.109)&0.750(0.058,~0.065)&1.000(0.000,~0.000)\\
&&$\hat A_{I1}$&0.497(0.067,~0.073)&0.621(0.061,~0.070)&0.730(0.053,~0.061)&0.862(0.034,~0.040)&1.000(0.000,~0.000)\\
&&$\hat A_{I2}$&0.495(0.065,~0.066)&0.622(0.061,~0.064)&0.729(0.053,~0.054)&0.859(0.029,~0.033)&0.980(0.008,~0.010)\\
&&$\hat A_{I3}$&0.496(0.065,~0.066)&0.622(0.062,~0.064)&0.729(0.051,~0.054)&0.858(0.030,~0.034)&0.983(0.004,~0.009)\\
&&$\hat \theta$&0.498(0.044,~0.049)&0.583(0.043,~0.049)&0.660(0.038,~0.044)&0.769(0.032,~0.036)&1.000(0.000,~0.004)\\ \\
%\cline{2-8}
100&(1.0,~1.0)&$\hat\rho$&0.085(0.056,~0.098)&0.253(0.083,~0.092)&0.497(0.082,~0.075)&0.747(0.046,~0.044)&1.000(0.000,~0.000)\\
&&$\hat A_{I1}$&0.490(0.050,~0.051)&0.620(0.043,~0.048)&0.739(0.046,~0.041)&0.865(0.024,~0.027)&1.000(0.000,~0.000)\\
&&$\hat A_{I2}$&0.485(0.053,~0.049)&0.622(0.041,~0.046)&0.741(0.042,~0.038)&0.864(0.023,~0.023)&0.987(0.007,~0.008)\\
&&$\hat A_{I3}$&0.483(0.054,~0.049)&0.620(0.042,~0.045)&0.739(0.042,~0.037)&0.861(0.024,~0.023)&0.982(0.006,~0.009)\\
&&$\hat \theta$&0.493(0.033,~0.034)&0.581(0.029,~0.033)&0.668(0.033,~0.030)&0.771(0.023,~0.024)&1.000(0.000,~0.001)\\ \\
%\cline{3-8}
&(1.0,~2.0)&$\hat\rho$&0.075(0.057,~0.097)&0.266(0.100,~0.091)&0.499(0.081,~0.074)&0.739(0.045,~0.046)&1.000(0.000,~0.000)\\
&&$\hat A_{I1}$&0.496(0.049,~0.051)&0.625(0.053,~0.048)&0.739(0.042,~0.041)&0.859(0.025,~0.027)&1.000(0.000,~0.000)\\
&&$\hat A_{I2}$&0.493(0.050,~0.049)&0.629(0.051,~0.045)&0.744(0.041,~0.037)&0.862(0.024,~0.023)&0.987(0.006,~0.006)\\
&&$\hat A_{I3}$&0.494(0.049,~0.049)&0.630(0.052,~0.045)&0.745(0.041,~0.037)&0.862(0.024,~0.024)&0.99(0.003,~0.005)\\
&&$\hat \theta$&0.498(0.032,~0.034)&0.586(0.036,~0.033)&0.667(0.031,~0.030)&0.765(0.022,~0.024)&1.000(0.000,~0.001)\\
\hline \multicolumn{8}{l}{$^{*}$Empirical standard deviations and
mean values of bootstrap standard deviations are in parentheses.}
\end{tabular*}}
\end{table*}
%\efloatseparator

Table \ref{tab2} shows the results of five measures when there is no
association between variable $Y$ and the gold standard $Z$. That is,
the data set used in this table are generated from the model
$y=z^2+\epsilon$ with standard normal error $\epsilon$, where the
gold standard $z$ is generated from three different distributions:
(1) normal distribution, (2) $t_2$ distribution with free degree 2,
and (3) a Cauchy distribution. Since $z$ has symmetrical density
functions for all three cases, it is clear that there is no
association between $Y$ and $Z$. That is, the ideal values of the
correlation coefficient estimate $|\hat\rho|$, ROC-type indexes
estimates $|\hat A_{Ij}-0.5|$,$j=1, 2, 3$ and $|\hat \theta-0.5|$
should be close to 0. We calculate the $25\%, 50\%$ and $75\%$
empirical quantiles based on 100 simulations.
The $p$-values, with a nominal significance level equal to 0.05, for
statistics $\hat \rho^2/V(\hat \rho)$, $(\hat A_{Ij}-0.5)^2/V(\hat
A_{Ij})$, $j=1, 2,3$ and $(\hat\theta-0.5)^2/V(\hat \theta)$ are
also reported.
It is seen from Table \ref{tab2} that all three quantiles of $\hat
A_{I3}$ and $\hat \theta$ are very close to 0, while the correlation
coefficient seems to over-estimate the association of $Y$ and $Z$ in
this experiment. When the tail of the distribution of $Z$ becomes
heavier, the quantiles and $p$-values of $\hat\rho$ become further
from 0.0 and nominal 0.05, respectively. Especially, when $Z$ is
from a Cauchy distribution, the 25\% quantiles are larger than 0.5
and the corresponding $p$-values are greater than 0.3.

The performances of $\hat A_{I3}$ and $\hat \theta$ are better than
those of $\hat A_{I1}$ and $\hat A_{I2}$ when $Z$ is not from a
normal distribution. This is because $\hat A_{I3}$ is based on a
kernel estimate of $f_c(t)$ and $\hat \theta$ is founded on a
nonparametric method, they are not affected by the distribution of
$Z$, and therefore very stable even when $Z$ is not normally
distributed.

As a summarization and conclusion to the results of Figure \ref{fig1},
and Tables \ref{tab1} and \ref{tab2},  both $\hat A_{I3}$ and $\hat
\theta$ are recommended for detecting the association between
variables and the continuous gold standard.
{Although $\hat\theta$ is considered as a natural extension of the
ordinary AUC index, it is worth noting that the performance of $\hat
A_{Ij}$ (especially $\hat A_{I3}$), in these cases, are  are very competitive.}

\begin{table*}[]
\tabcolsep=3pt \fontsize{9}{12}\selectfont
\begin{center}
\begin{minipage}{160mm}
\begin{center}
\caption{Comparison of different methods when there is no
association between variable $Y$ and the gold standard $Z$. The data
set $(y,z)$ is generated from model $y=z^2+\epsilon$ with standard
normal error $\epsilon$. Three different distributions of $z$ are
used, which are a normal distribution, a $t_2$ distribution with
free degree 2 and a Cauchy distribution.} \label{tab2}\vspace{10pt}
\begin{tabular*}{\textwidth}{@{}c@{\extracolsep{\fill}}c@{\extracolsep{\fill}}c@{\extracolsep{\fill}}c
 @{\extracolsep{\fill}}c@{\extracolsep{\fill}}c@{\extracolsep{\fill}}c@{\extracolsep{\fill}}c@{\extracolsep{\fill}}c@{\extracolsep{\fill}}c}
 \hline
 % after \\: \hline or \cline{col1-col2} \cline{col3-col4} ...
 & &\multicolumn{2}{c}{ Normal}&& \multicolumn{2}{c}{ $t_2$}&& \multicolumn{2}{c}{ Cauchy}\\
 \cline{3-4}   \cline{6-7}   \cline{9-10}
$n$&Model &$(25\%,~50\%,~75\%)$& p-value$^{*}$&& $(25\%,~50\%,~75\%)$& p-value&& $(25\%,~50\%,~75\%)$& p-value\\
 %\hline

50&$\hat\rho$&(0.086, 0.175, 0.287)&0.13&&(0.277, 0.544, 0.802)&0.30&&(0.515, 0.834, 0.948)&0.33\\
&$\hat A_{I1}$&(0.031, 0.060, 0.110)&0.09&&(0.048, 0.079, 0.125)&0.12&&(0.044, 0.074, 0.125)&0.13\\
&$\hat A_{I2}$&(0.025, 0.054, 0.090)&0.06&&(0.041, 0.076, 0.120)&0.13&&(0.040, 0.078, 0.150)&0.15\\
&$\hat A_{I3}$&(0.022, 0.050, 0.087)&0.06&&(0.036, 0.060, 0.090)&0.09&&(0.028, 0.055, 0.088)&0.09\\
&$\hat \theta$&(0.021, 0.043, 0.081)&0.08&&(0.034, 0.060, 0.087)&0.09&&(0.025, 0.049, 0.090)&0.08\\  \\
%\cline{2-10}
100&$\hat\rho$&(0.055, 0.114, 0.202)&0.07&&(0.305, 0.527, 0.728)&0.27&&(0.603, 0.825, 0.931)&0.37\\
&$\hat A_{I1}$&(0.022, 0.044, 0.072)&0.07&&(0.016, 0.035, 0.072)&0.06&&(0.029, 0.052, 0.098)&0.15\\
&$\hat A_{I2}$&(0.020, 0.033, 0.059)&0.06&&(0.014, 0.033, 0.064)&0.05&&(0.026, 0.048, 0.096)&0.14\\
&$\hat A_{I3}$&(0.017, 0.034, 0.060)&0.06&&(0.009, 0.032, 0.056)&0.04&&(0.018, 0.041, 0.07)&0.08\\
&$\hat \theta$&(0.015, 0.032, 0.049)&0.07&&(0.012, 0.031, 0.054)&0.04&&(0.014, 0.036, 0.065)&0.06\\

\hline \multicolumn{10}{l}{$^{*}${Nominal significance level is
0.05}.}
\end{tabular*}
\end{center}
\end{minipage}
\end{center}
\end{table*}

%\efloatseparator

\begin{table*}[]
\tabcolsep=3pt \fontsize{9}{12}\selectfont
\begin{center}
\begin{minipage}{130mm}
\begin{center}
\caption{Results of linear combination using correlation
coefficient (CC) and TGDM method.}\label{tab3} %\vspace{-5pt}
%by using slope piecewise constant $H(t)$

\begin{tabular*}{\textwidth}{@{}c@{\extracolsep{\fill}}c@{\extracolsep{\fill}}c@{\extracolsep{\fill}}c
 @{\extracolsep{\fill}}c@{\extracolsep{\fill}}c@{\extracolsep{\fill}}c@{\extracolsep{\fill}}c}
\hline
& & & &\multicolumn{2}{c}{ $Nonzero~coef.^{+}$}& &\\
 \cline{5-6}
Distribution&$p^{**}$&$n$ &Method & $x_1$& $x_2$ &CC&TGDM\\

% \hline
$Normal$&4&50&$\hat A_{I3}$&0.773(0.054)$^{*}$&0.786(0.052)&0.900(0.024)&0.900(0.028)\\
&&&$\hat \theta$&0.694(0.043)&0.702(0.042)&0.815(0.028)&0.815(0.031)\\ \\
&&100&$\hat A_{I3}$&0.782(0.033)&0.785(0.035)&0.904(0.018)&0.906(0.018)\\
&&&$\hat \theta$&0.693(0.027)&0.696(0.030)&0.807(0.021)&0.809(0.021)\\ \\
&10&50&$\hat A_{I3}$&0.785(0.048)&0.773(0.046)&0.909(0.021)&0.900(0.031)\\
&&&$\hat \theta$&0.703(0.037)&0.692(0.040)&0.824(0.027)&0.815(0.033)\\
&&100&$\hat A_{I3}$&0.791(0.036)&0.789(0.032)&0.913(0.015)&0.913(0.016)\\
&&&$\hat \theta$&0.699(0.030)&0.700(0.025)&0.818(0.019)&0.817(0.020)\\ \\
&20&50&$\hat A_{I3}$&0.767(0.051)&0.779(0.053)&0.928(0.018)&0.897(0.034)\\
&&&$\hat \theta$&0.689(0.042)&0.698(0.042)&0.852(0.026)&0.813(0.039)\\
&&100&$\hat A_{I3}$&0.782(0.033)&0.783(0.032)&0.922(0.015)&0.915(0.016)\\
&&&$\hat \theta$&0.693(0.028)&0.696(0.025)&0.828(0.019)&0.820(0.019)\\
\\

$Cauchy$&4&50&$\hat A_{I3}$&0.659(0.067)&0.640(0.068)&0.669(0.107)&0.735(0.073)\\
&&&$\hat \theta$&0.629(0.046)&0.614(0.046)&0.619(0.088)&0.685(0.059)\\ \\
&&100&$\hat A_{I3}$&0.660(0.056)&0.657(0.047)&0.659(0.094)&0.724(0.077)\\
&&&$\hat \theta$&0.629(0.036)&0.625(0.032)&0.615(0.078)&0.679(0.063)\\ \\
&10&50&$\hat A_{I3}$&0.648(0.064)&0.645(0.072)&0.690(0.099)&0.750(0.067)\\
&&&$\hat \theta$&0.620(0.045)&0.618(0.048)&0.628(0.079)&0.689(0.056)\\
&&100&$\hat A_{I3}$&0.648(0.083)&0.638(0.082)&0.664(0.104)&0.733(0.101)\\
&&&$\hat \theta$&0.625(0.033)&0.618(0.035)&0.614(0.063)&0.683(0.061)\\ \\
&20&50&$\hat A_{I3}$&0.647(0.093)&0.657(0.096)&0.740(0.123)&0.789(0.096)\\
&&&$\hat \theta$&0.623(0.044)&0.628(0.046)&0.665(0.083)&0.719(0.052)\\
&&100&$\hat A_{I3}$&0.634(0.123)&0.638(0.120)&0.649(0.142)&0.739(0.147)\\
&&&$\hat \theta$&0.624(0.032)&0.627(0.029)&0.604(0.068)&0.689(0.069)\\

\hline
%\multicolumn{12}{l}{$\hat A_{I3}$ denotes the smoothed estimate of $AUC_{Is}$ with the kernel density estimate of weight function $f_c(t)$.}\\
\multicolumn{8}{l}{$^{+}${$Nonzero~coef.$ represents variables with non-zero coefficients in true model}.}\\
\multicolumn{8}{l}{$^{*}${Empirical standard deviations are in parentheses}.}\\
\multicolumn{8}{l}{$^{**}${$p$ denotes number of total variables in
true model and the number of}}\\
\multicolumn{8}{l}{{~~~~non-zero variables is $p_1=2$}.}
\end{tabular*}
\end{center}
\end{minipage}
\end{center}
\end{table*}
%\efloatseparator

\subsection{Combination of Variables}
%We also investigate the ability of finding the linear combination of variables.
Both correlation coefficient (CC) and the TGDM algorithm are used to
obtain the optimal linear combinations of variables. We then
calculate $\hat A_{I3}$ and $\hat \theta$ of the corresponding
combination of variables based on the coefficient vectors obtained
from these two methods. The threshold parameter $\tau$ in the TGDM
algorithm is equal to 1.0 in our studies. The data set are generated
from $Z=l^{T}\tilde{X}+\epsilon$, where $\tilde{X}$ follows a $p$
dimensional multivariate normal distribution with mean vector $(0,
\ldots, 0)^T$ and an identity covariance matrix, and the true
$l=(1.0,1.0,0.0,\cdots,0.0)^{T}$. Error term $\epsilon$ is generated
from either the standard normal distribution or a Cauchy
distribution. In this experimental study, we have tried three
different dimensions of $X$ ($p=4, 10, 20$) for all cases, and only
variables $x_1$ and $x_2$ have non-zero coefficients. That is, only
these two variables are associated with the gold standard. Moreover,
a software based on the TGDM algorithm to calculate the optimal
linear combination of variables is available as an R package. It is
also worth noting that there is no algorithm or discussion in
\cite{Obuchowski2005} about finding the linear combination of
variables based on $\hat\theta$.

Table \ref{tab3} lists the values of $\hat A_{I3}$ and $\hat \theta$
for individual variables, $x_1$ and $x_2$, and the linear
combinations based on the CC and TGDM methods.
From this table, we find that $\hat A_{I3}$ and $\hat \theta$ for
linear combinations of variables are always larger than for
individual variables, which confirms that linear combinations of
variables can improve on the the diagnostic power of individual
variables.
When $\epsilon$ follows the standard normal distribution, $\hat
A_{I3}$ and $\hat \theta$ for linear combinations based on both TGDM
and CC are very close. However, when $\epsilon$ is a Cauchy
distribution, the TGDM method has larger $\hat A_{I3}$ and $\hat
\theta$ than combinations based on CC.  This is because the  CC
method relies on the normality assumption, while TGDM does not. In
addition, from Table \ref{tab3}, we can see that $\hat A_{I3}$ is
larger than $\hat \theta$. In most of the cases, the standard
deviations of TGDM are smaller than those of $\hat \theta$, which
suggests that the linear combinations based on TGDM have greater
diagnostic power, although the difference may not be statistically
significant in our simulation.

%\vskip 1cm
\subsection{Real examples}

We apply the proposed measures to three real data sets: tumor,
prostate and diabetes data sets, which are used in
\cite{Obuchowski2005}, \cite{Stamey1989} and \cite{Willems-etal97},
respectively. In the tumor data set, there are 74 patients and only
two surgery variables: the computed tomography (CT) and a fictitious
test (Fi). The continuous gold standard of this data set is the size
of the renal tumor mass. The prostate data has 97 patients with
prostate specific antigen as its gold standard together with 6
continuous variables, which are cancer volume, prostate weight, age
(Age), benign prostatic hyperplasia amount, capsular penetration,
and percentage Gleason scores 4 or 5 (Pgg45). Except variables Age
and Pgg45, the others are re-coded in log-scale and denoted by
Lcavol, Lweight, Lbph, Lcp and Lpsa, accordingly. The original
diabetes data consists of 403 subjects, but we follows
\cite{Willems-etal97} to delete 22 subjects  with missing variables.
Of the remaining 381 subjects from this data set used in our
numerical study, 222 are females and 159 are males. The following 8
continuous variables are used in this data set: total cholesterol
(Chol), stabilized glucose (Stab.glu), high density lipoprotein
(Hdl), cholesterol/HDL ratio (Ratio), age (Age), body mass index
(BMI) and waist/hip ratio (WHR). The gold standard for this data set
is glycosylated hemoglobin (Glyhb), which is commonly used as a
measure of the progress of diabetes. In addition to analyzing the
entire diabetes data set,  we also investigate female and male
subgroups, separately.

We normalize the data  before applying the proposed measures to each
data set to avoid  scale variations. Table \ref{tab4} presnets $\hat
A_{I3}$ and $\hat \theta$ for individual variables with $p$-value
less than $10^{-7}$.  From Table \ref{tab4}, we find that $\hat
A_{I3}$ selects more variables than $\hat \theta$ for some cases.
Note that $\hat A_{I3}$ are much larger than $\hat \theta$
with competitive standard deviations in these cases.

Table \ref{tab5} lists the linear coefficients obtained using the
TGDM and  CC methods, and their corresponding $\hat A_{I3}$ and
$\hat \theta$ values for all data sets, including the male and
female subgroups of the diabetes data set. In the tumor data set, Fi
has a larger $\hat A_{I3}$ value than CT; that is, Fi has a greater
association with the size of the renal tumor mass for tumor data. In
the prostate data set, Lcavol has the largest $\hat A_{I3}$ value;
that is, Lcavol is most highly associated with prostate specific
antigen among all variables considered in the prostate data set. For
the diabetes data set and its male and female subgroups, the largest
$\hat A_{I3}$ and the variable with the largest coefficient value is
Stab.glu; that is, Stab.glu has the highest potential to diagnose
diabetes in terms of glycosylated hemoglobin index.
As expected, from Tables \ref{tab4} and \ref{tab5}, the linear
combinations based on TGDM and CC usually have larger $\hat A_{I3}$
and $\hat \theta$ values than individual variables do, and
similarly, $\hat A_{I3}$ and $\hat \theta$ values for combinations
from TGDM are a little bit larger than those obtained using the CC
method.  In real data sets the relation is seldom linear, which is
the reason why the combinations obtained using TGDM perform better
than others.

\begin{table*}[]
\tabcolsep=1.5pt \fontsize{9}{12}\selectfont
\begin{center}
\begin{minipage}{120mm}
\begin{center}
\caption{Results of ROC measure indexes: $\hat A_{I3}$ and
$\hat\theta$, of single markers for tumor, prostate, diabetes,
diabetes-female and diabetes-male data sets.}\label{tab4}
\vspace{10pt} Tumor
\vspace{10pt}\\
\begin{tabular*}{\textwidth}{@{}c@{\extracolsep{\fill}}c@{\extracolsep{\fill}}c@{\extracolsep{\fill}}c}

 \hline
 % after \\: \hline or \cline{col1-col2} \cline{col3-col4} ...
Data&Method &  CT & Fi \\
Tumor&$\hat A_{I3}$&0.943(0.014)$^{*}$&0.982(0.011)\\
&$\hat\theta$&0.871(0.020)&0.956(0.008)\\
\hline\\
\\
\end{tabular*}
\vspace{10pt}
Prostate
\vspace{10pt}
\begin{tabular*}{\textwidth}{@{}c@{\extracolsep{\fill}}c@{\extracolsep{\fill}}c@{\extracolsep{\fill}}c
@{\extracolsep{\fill}}c@{\extracolsep{\fill}}c}
 \hline
 % after \\: \hline or \cline{col1-col2} \cline{col3-col4} ...
Data&Method &  Lcavol & Lweight&Lcp&Pgg45\\
Prostate&$\hat A_{I3}$&0.865(0.022)&0.722(0.034)&0.759(0.035)&0.744(0.035)\\
&$\hat\theta$&0.758(0.027)&0.647(0.027)&0.675(0.031)&0.676(0.028)\\
\hline\\
\\
\end{tabular*}
\vspace{10pt}
Diabetes
\vspace{10pt}
\begin{tabular*}{\textwidth}{@{}c@{\extracolsep{\fill}}c@{\extracolsep{\fill}}c@{\extracolsep{\fill}}c
@{\extracolsep{\fill}}c@{\extracolsep{\fill}}c}
\hline
 Data&Method & Chol & Stab.glu & Ratio & Age \\

Diabetes&$\hat A_{I3}$&-&0.779(0.021)&0.662(0.022)&0.711(0.019)\\
&$\hat\theta$&-&0.687(0.017)&0.600(0.015)&0.644(0.014)\\
\\
%\cline{2-6}
Diabetes-&$\hat A_{I3}$&0.667(0.029)&0.786(0.022)&-&0.732(0.025)\\
female&$\hat\theta$&-&0.691(0.021)&-&0.665(0.019)\\
\\

%\cline{2-6}
Diabetes-&$\hat A_{I3}$&-&0.769(0.039)&0.689(0.034)&0.681(0.030)\\
male&$\hat\theta$&-&0.682(0.030)&-&-\\

\hline
\multicolumn{6}{l}{$^{*}${Bootstrap standard deviation is in
parentheses}.}
\end{tabular*}
\end{center}
\end{minipage}
\end{center}
\end{table*}

\begin{table*}[]
\tabcolsep=3pt \fontsize{8}{10}\selectfont
\begin{center}
\begin{minipage}{150mm}
\begin{center}
\caption{Results of optimal linear coefficients and corresponding
ROC measure indexes: $\hat A_{I3}$ and $\hat\theta$, for tumor,
prostate, diabetes, diabetes-female and diabetes-male data sets.}
\label{tab5}
\vspace{10pt}
Tumor
\vspace{10pt}\\

\begin{tabular*}{\textwidth}{@{}c@{\extracolsep{\fill}}c@{\extracolsep{\fill}}c@{\extracolsep{\fill}}c
@{\extracolsep{\fill}}c@{\extracolsep{\fill}}c@{\extracolsep{\fill}}c}
\hline
 Data&Method & \multicolumn{2}{c}{Coef.} & &\multicolumn{2}{c}{ROC-type indexes}\\
\cline{3-4}  \cline{6-7}
 && CT &Fi&&$\hat A_{I3}$&$\hat\theta$\\
 Tumor&CC&-0.118&1.076&&0.981(0.011)&0.950(0.009) \\
&TGDM&0.044&1.000&&0.983(0.011)&0.957(0.008)\\
\hline\\
\\
\end{tabular*}
\vspace{10pt}
Prostate
\vspace{10pt}
\begin{tabular*}{\textwidth}{@{}c@{\extracolsep{\fill}}c@{\extracolsep{\fill}}c@{\extracolsep{\fill}}c
@{\extracolsep{\fill}}c@{\extracolsep{\fill}}c@{\extracolsep{\fill}}c@{\extracolsep{\fill}}c
@{\extracolsep{\fill}}c@{\extracolsep{\fill}}c@{\extracolsep{\fill}}c}
\hline
 Data&Method & \multicolumn{6}{c}{Coef.} & &\multicolumn{2}{c}{ROC-type indexes}\\
\cline{3-8}  \cline{10-11}
&& Lcavol & Lweight& Age&Lbph&Lcp&Pgg45&&$\hat A_{I3}$&$\hat\theta$\\
Prostate&CC&0.642&0.214&-0.118&0.099&0.017&0.147&&0.892(0.018)&0.791(0.024)\\
&TGDM&1.000&0.264&-0.108&0.135&-0.013&0.189&&0.891(0.017)&0.789(0.023)\\
\hline\\
\\
\end{tabular*}
\vspace{10pt}
Diabetes
\vspace{10pt}
\begin{tabular*}{\textwidth}{@{}c@{\extracolsep{\fill}}c@{\extracolsep{\fill}}c@{\extracolsep{\fill}}c
@{\extracolsep{\fill}}c@{\extracolsep{\fill}}c@{\extracolsep{\fill}}c@{\extracolsep{\fill}}c
@{\extracolsep{\fill}}c@{\extracolsep{\fill}}c@{\extracolsep{\fill}}c@{\extracolsep{\fill}}c}
\hline
 Data&Method & \multicolumn{7}{c}{Coef.} & &\multicolumn{2}{c}{ROC-type indexes}\\
\cline{3-9}  \cline{11-12}
&& Chol & Stab.glu & Hdl & Ratio & Age & BMI&WHR&&$\hat A_{I3}$&$\hat\theta$\\
Diabetes&CC&0.074&0.668&0.018&0.101&0.101&0.017&0.019&&0.816(0.017)&0.717(0.015)\\
&TGDM&0.061&1.000&-0.027&0.099&0.373&0.140&0.011&&0.826(0.018)&0.723(0.016)\\
\\
%\cline{2-12}
Diabetes-female&CC&0.109&0.659&-0.073&0.027&0.106&0.029&0.069&&0.834(0.021)&0.737(0.019)\\
&TGDM&0.253&1.000&-0.164&-0.007&0.389&0.133&0.199&&0.842(0.019)&0.741(0.018)\\
\\
%\cline{2-12}
Diabetes-male&CC&-0.005&0.701&0.141&0.243&0.085&-0.049&-0.002&&0.786(0.03)&0.691(0.025)\\
&TGDM&-0.016&1.000&0.009&0.179&0.367&0.100&-0.040&&0.811(0.031)&0.706(0.027)\\
\hline
\multicolumn{12}{l}{$^{*}${ROC-type indexes used here are $AUC_{I3}$ and
$\hat\theta$}.}
\end{tabular*}
\end{center}
\end{minipage}
\end{center}

\end{table*}

\section{Conclusion and Discussion}
In this paper, we first propose a new measure for evaluating the potential
diagnostic power of individual variables, when there is only a
continuous gold standard available and no confirmative threshold for
it is known.  The proposed measure is an AUC-type index that shares
the threshold independent property of the ROC curve and AUC, and can also be used to evaluate the performance of classifiers when the gold standard variable is essentially continuous, and the threshold is controvertible.
Numerical results show that the proposed novel index is very competitive to the existence method.

{In addition, we propose algorithms, based on the newly defined index, for finding the best linear combination of variables, which is useful from a practical prospect when there are multiple variables considered at a time, and how to evaluate or select a good combination of variables is an important issue.}
Here we also study numerical methods for finding the linear combination of variables that maximizes the
proposed measure.  When the normality assumption of variables is
valid, the best linear combination solution can be realized as a
solution to a linear system.  Thus, under an assumption of normality
and when the number of variable $p$ is large, the LARS algorithm can
be applied to obtain such a linear combination. This also implies
that  the LARS-type variable selection scheme can be conducted even
when no binary-scale gold standard is available.
When the joint distribution of variables is unknown, the proposed
measure is then approximated using a nonparametric kernel density
estimation method. In this case, we proposed a TGDM-based algorithm
to calculate the best linear combination of variables.  Based on
numerical results, we found that our method is numerically stable
with computational advantage when there are large number of variables
considered and combination of variables is of interest.
Moreover, our method can be easily extended to an ordinal-scale gold standard with
a suitable choice of a weight function for cutting points, which will be reported elsewhere.

\section*{Appendix}

Let random variables $(Y_i, Z_i)$ denote a pair of measures from
subject $i$, for $i \geq 1$. Suppose that
$\{(y_i,z_i),~i=1,\ldots,n\}$ are $n$ independent observed values of
random variables $(Y_i, Z_i)$, $i=1, \cdots, n$. For a given cutting
point $c$, a subject $i$, $i=1, \ldots, n$, is assigned as a
``case'' if $z_i>c$ and otherwise labeled as a ``control''. That is,
for a given $c$, we divide the observed subjects into two groups;
let $S_1(c)$ and $S_0(c)$ be the case and control groups with sample
sizes $n_1$ and $n_0$, respectively.

Then we propose a natural estimate of AUC index, $AUC_I$, with
continuous gold standard,
\begin{eqnarray}\label{gauc_est}
    \hat A_{I}=\int \hat A(t) d \hat F_c(t),
\end{eqnarray}
where $\hat A(c)$ is defined as
\begin{eqnarray}\label{AUChat}
    \hat A(c)=\frac{1}{n_0 \, n_1}\sum_{i\in S_1(c); \, j\in S_0(c)} \psi(y_i -
    y_j),\nonumber
\end{eqnarray}
$\psi(u)=1$, if $u>0$; $=0.5$, if $u=0$ and $=0$ if $u<0$ and $\hat
F_c(t)$ is the empirical estimate of the cumulative distribution
function of $Z$ based on $\{z_1, \ldots, z_n\}$. However, in
practice, it is rare to choose cutting points at ranges near the two
ends of the distribution of $Z$.  Thus, instead of the whole range
of $Z$, we might explicitly define a weight function $f_c(t)$ on a
particular critical range.

Since the step function $\psi(\cdot)$ in (\ref{gauc_est}) is not
continuously differentiable, a smooth estimate of $AUC_I$ is defined
as
\begin{eqnarray}\label{asauc_I}
    \hat A_{Is}=\int \frac{1}{n_1 n_0}\sum_{i\in S_1(t);j\in
    S_0(t)}S\left (\frac{y_i-y_j}{h}\right )f_c(t) dt,
\end{eqnarray}
where $S(t)$ is a sigmoid function $1/(1+\exp(-t))$ and $h$ is
window width.

\section*{Appendix A: Proof of Strong Consistency of $\hat A_{Is}(l)$}
%For convenience, we let $C$ denote some positive constant independent of $n$, which may take different values in different formulae or even in different parts of one and the same inequality.

The proof of the strong consistency of smoothed $AUC_I(l)$ estimator
$\hat A_{Is}(l)$ follows from the following three lemmas.
%\vskip 10pt
%\noindent{\textbf{Lemma A.1}}:
\begin{lemma}\label{lemma:1}
 Suppose that $X_1,\cdots,X_n$ is a sequence of independent
and identically distributed random variables with values in $R^1$,
and a uniformly continuous density $f(\cdot)$. Let $k(x)$ be a
bounded probability density and the Dirichlet series
$\sum_{n=1}^{\infty}n \, \exp({-\gamma \eta_n})$, $\eta_n=n \,
h^{2}$ converges for any $\gamma>0$. Then
$$\int_{-\infty}^{\infty}|f_n(x)-f(x)|dx\rightarrow 0,
~{almost~surely}\hbox{~as~} n\rightarrow \infty,$$ where
$f_n(x)=\frac{1}{n \, h}\sum_{i=1}^{n}k((x-X_i)/h)$ is a kernel
density estimator of $f(x)$.
\end{lemma}
(The proof of Lemma \ref{lemma:1} can be found in
\cite{Nadaraya:1989}, Theorem 3.1, page 55. So, it is omitted here.)

%\vskip 10pt
%\noindent{\textbf{Lemma A.2}}:
\begin{lemma} \label{lemma:2}
  Suppose that
$X_1,\cdots,X_n$ is a sequence of independent and identically
distributed random variables with values in $R^1$, and a uniformly
continuous density. Then with probability one, as $n\rightarrow
\infty$
$$\sup_{x\in R^1}|F_n(x)-F(x)|\rightarrow 0,$$
where $F_n(\cdot)$ and $F(\cdot)$ are the empirical distribution and
distribution functions of $X$, respectively.
\end{lemma}

%\begin{proof}[ of Lemma \ref{lemma:2}]
\noindent Proof of Lemma \ref{lemma:2}:\\
From \cite{Nadaraya:1989} (Equation (1.4), page 43), we have
\begin{eqnarray}\label{ineq1-2}
{\rm{pr}}(\sup_{x\in R^1}|F_n(x)-F(x)|>\eta \, n^{-1/2})\le c \,
\exp({-2\eta^2}),
\end{eqnarray}
which completes the proof of Lemma \ref{lemma:2}.
%\end{proof}

%\vskip 10pt
%\noindent{\textbf{Lemma A.3}}:
\begin{lemma} \label{lemma:3}
Assume that $\{(y_1,~z_1),\cdots,(y_n,~z_n)\}$ are $n$ independent
and identically distributed samples of $(Y \in R^1, Z\in R^1)$,
where $Z$ denotes a continuous gold standard. For a given $c$, let
$\tilde{f}(y|Z>c)$ be a conditional density function of $Y$ given
$Z>c$. Suppose that conditions of Theorem 3 %\ref{thm3-gauc}
holds. Then $\tilde{f}(\cdot|Z>c)$ is uniformly continuous.
\end{lemma}

%\vskip 10pt
%\noindent
%\begin{proof}[ of Lemma \ref{lemma:3}]
\noindent Proof of Lemma \ref{lemma:3}:\\
By the Bayesian theorem, we have
\begin{eqnarray}\label{ineq6}
\tilde{f}(y|Z>c)=\frac{\int_{c}^{\infty}f(y,z)dz}{{\rm{pr}}(Z>c)}.
\end{eqnarray}
For any $y_i\in R^{1}$, $i=1,2$,
\begin{eqnarray}\label{ineq7}
&\int_{c}^{\infty}f(y_1,z)dz-\int_{c}^{\infty}f(y_2,z)dz\nonumber\\
&=\int_{c}^{\infty}[f(z|y_1)f_{Y}(y_1)-f(z|y_1)f_{Y}(y_2)]dz
+\int_{c}^{\infty}[f(z|y_1)f_{Y}(y_2)-f(z|y_2)f_{Y}(y_2)]dz\nonumber\\
&=[f_{Y}(y_1)-f_{Y}(y_2)][1-F(c|y_1)]+[F(z|y_2)-F(z|y_1)]
f_{Y}(y_2),
\end{eqnarray}
where $f(z|y)$ is a conditional density function of $Z$ given $Y=y$
and $f_{Y}(y)$ is a density function of marker $Y$. From the
conditions of Theorem 3, we have $b\equiv {\rm{pr}}(Z>c)>0$,
$f_{Y}(\cdot)<M$ and both $f_{Y}(\cdot)$ and $F(z|\cdot)-F(z|\cdot)$
are uniformly continuous. Hence, for any $\epsilon>0$, there exists
a $\delta>0$, for any $y_1$ and $y_2$ satisfying $|y_1-y_2|<\delta$,
we have
\begin{eqnarray}\label{ineq7-1}
&|f_{Y}(y_1)-f_{Y}(y_2)|<b\epsilon/2\nonumber\\
&|F(z|y_2)-F(z|y_1)|<b\epsilon/(2M).
\end{eqnarray}
Consequently, by (\ref{ineq6}), (\ref{ineq7}) and (\ref{ineq7-1}) we
get that for a given $c$,
\begin{eqnarray}\label{ineq7-2}
& &|\tilde{f}(y_1|Z>c)-\tilde{f}(y_2|Z>c)|\nonumber\\
&<&\frac{1}{b}\{|f_{Y}(y_1)-f_{Y}(y_2)|(1-F(c|y_1))+|F(z|y_2)-F(z|y_1)|f_{Y}(y_2)\}\nonumber\\
&<&\epsilon/2+\epsilon/2=\epsilon.
\end{eqnarray}
It follows that $\tilde{f}(\cdot|Z>c)$ is uniformly continuous.
%\end{proof}

\vskip 10pt
%\noindent
%\begin{proof}[ of Theorem \ref{thm3-gauc}.] %$\hat A_{Is}(l)$ strong consistency}:\\
\noindent Proof of Theorem 3:\\
By the triangle inequality, we have, for fixed $l$,
\begin{eqnarray}\label{ineq1}
\left|\hat A_{Is}-AUC_I\right|
&\leq \left|\hat A_{Is}-\hat A_I\right|+\left|\hat A_I-{AUC}_I\right|\nonumber \\
&= (I) +(II) \hbox{~(say)}.
\end{eqnarray}
From Theorem 1, (II) converges to 0 almost surely as $n$ goes to
$\infty$; that is
\begin{eqnarray}\label{ineq1-0}
\hat A_{I}-{AUC}_I \rightarrow 0  \, \hbox{~~~~almost~surely}
\hbox{~as~} n \rightarrow \infty.
\end{eqnarray}

%Now let us prove that the first term of (\ref{ineq1}) almost surely converges to 0 as $n$ tends to $\infty$.
From (\ref{gauc_est})  and (\ref{sauc_I}),
\begin{eqnarray}
%&  = \left|\hat A_{Is}(l)-\hat A_I(l)\right|\nonumber\\
&(I)= \left|\int\!\! \frac{1}{n_1 n_0}\sum_{i\in S_1(t);j\in S_0(t)}
\!S\left (\frac{y_i-y_j}{h}\right )f_c(t) dt\!-\!\int \frac{1}{n_1
n_0}\sum_{i\in S_1(t);j\in S_0(t)}\!\psi(y_i-y_j) f_c(t) dt\right|
\nonumber\\
&\leq \int \left|\frac{1}{n_1 n_0}\sum_{i\in S_1(t);j\in
S_0(t)}S\left (\frac{y_i-y_j}{h}\right )- \frac{1}{n_1
n_0}\sum_{i\in S_1(t);j\in S_0(t)}\psi(y_i-y_j)\right| f_c(t)
dt.\nonumber
\end{eqnarray}
Due to $n_1+n_0=n$, then at least one of $n_1\rightarrow \infty$ and
$n_0\rightarrow \infty$ holds as $n$ tends to $\infty$. Without loss
of generality, assume that $n_1$ tends to $\infty$. Then
\begin{eqnarray}\label{ineq2}
%& \left|\hat A_{Is}(l)-\hat A_I(l)\right|\nonumber\\
&(I) \leq \int \frac{1}{n_0}\sum_{j\in S_0(t)}
\left|\frac{1}{n_1}\sum_{i\in S_1(t)}S\left (\frac{y_i-y_j}{h}\right
)
-\tilde{F}(y_j|Z>t)\right| f_c(t) dt\nonumber\\
&\hbox{~~~}+\int \frac{1}{n_0}\sum_{j\in S_0(t)}
\left|\frac{1}{n_1}\sum_{i\in S_1(t)}\psi(y_i-y_j)
-\tilde{F}(y_j|Z>t)\right| f_c(t) dt,
\end{eqnarray}
where $\tilde{F}(\cdot|Z>t)$ is the conditional cumulative
distribution function of $Y$ given $\{Z>t\}$.
%Corresponding to $\tilde{F}(\cdot|Z>c)$,
Let $\tilde{f}(\cdot|Z>t)$ be its conditional density function. By
Lemma \ref{lemma:3}, $\tilde{f}(\cdot|Z>t)$ is uniformly continuous.

Let $h=n^{-\alpha}$, $1/5<\alpha<1/2$.
%(The $h$ here is not directly connected to other parts of proof. Which lemma is this condition for? It needs to be clarify.)}
Set $\eta_n=nh^2=n^{1-2\alpha}$, and the Dirichlet series
$\sum_{n=1}^{\infty}n \exp({-\gamma \eta_n})$ converges for any
$\gamma>0$. Thus, the conditions of Lemma \ref{lemma:1} are
satisfied. Let $k(t)$ denote the derivative of $S(t)$, then $k(t)$
is a bounded probability density. Thus, by Lemma \ref{lemma:1},
\begin{eqnarray}\label{ineq3}
& \sup_{y\in R^1}\left|\frac{1}{n_1}\sum_{i\in S_1(t)}S\left
(\frac{y_i-y}{h}\right )
-\tilde{F}(y|Z>t)\right|\nonumber\\
&=\sup_{y\in
R^1}\left|\int_{-\infty}^{y}\left(\frac{1}{n_1h}\sum_{i\in
S_1(t)}k\left (\frac{y_i-t}{h}\right )
-\tilde{f}(t|Z>t)\right)dt\right|\nonumber\\
&\leq \int_{-\infty}^{\infty}\left|\left(\frac{1}{n_1h}\sum_{i\in
S_1(t)}k\left (\frac{y_i-t}{h}\right )
-\tilde{f}(t|Z>t)\right)\right|dt \longrightarrow 0,
\hbox{~~~~almost~surely} \hbox{~as~} n\rightarrow \infty.
\end{eqnarray}
From Lemma \ref{lemma:2}, we have
\begin{eqnarray}\label{ineq4}
\sup_{y\in R^1}\left|\frac{1}{n_1}\sum_{i\in S_1(t)}\psi(y_i-y)
-\tilde{F}(y|Z>t)\right| \longrightarrow 0, \hbox{~~~~almost~surely}
\hbox{~as~} n\rightarrow \infty.
\end{eqnarray}
From (\ref{ineq2}), (\ref{ineq3}) and (\ref{ineq4}), we prove that
\begin{eqnarray}\label{ineq5}
\hat A_{Is}-\hat A_I \rightarrow 0, \hbox{~almost~surely}
\hbox{~as~} n \rightarrow \infty.
\end{eqnarray}
Put (\ref{ineq1-0}) and (\ref{ineq5}) together to complete the proof
of Theorem 3.
%\end{proof}

\section*{Acknowledgements}

This work is partially supported via NSC97-2118-M-001-004-MY2 funded by the National Science Council,
Taipei, Taiwan, ROC.
\vspace*{-8pt}

\bibliographystyle{elsarticle-harv}
\bibliography{<your-bib-database>}

%% Authors are advised to submit their bibtex database files. They are
%% requested to list a bibtex style file in the manuscript if they do
%% not want to use elsarticle-harv.bst.

%% References without bibTeX database:

% \begin{thebibliography}{00}

%% \bibitem must have one of the following forms:
%%   \bibitem[Jones et al.(1990)]{key}...
%%   \bibitem[Jones et al.(1990)Jones, Baker, and Williams]{key}...
%%   \bibitem[Jones et al., 1990]{key}...
%%   \bibitem[\protect\citeauthoryear{Jones, Baker, and Williams}{Jones
%%       et al.}{1990}]{key}...
%%   \bibitem[\protect\citeauthoryear{Jones et al.}{1990}]{key}...
%%   \bibitem[\protect\astroncite{Jones et al.}{1990}]{key}...
%%   \bibitem[\protect\citename{Jones et al., }1990]{key}...
%%   \harvarditem[Jones et al.]{Jones, Baker, and Williams}{1990}{key}...
%%

% \bibitem[ ()]{}

% \end{thebibliography}

\end{document}